\documentclass[reprint,superscriptaddress,nofootinbib,amsmath,amssymb,aps]{revtex4-2}

\usepackage{graphicx}
\usepackage{dcolumn}
\usepackage{bm}
\usepackage{braket}
\usepackage{mathrsfs}
\usepackage{hyperref}
\hypersetup{
    colorlinks,%
    citecolor=blue,%
    filecolor=blue,%
    linkcolor=blue,%
   	urlcolor=blue,
   	linktoc=page
}
\allowdisplaybreaks

\usepackage{mathtools}
\usepackage{mathrsfs}
\usepackage{setspace}
\usepackage{tikz}
\usetikzlibrary{calc}
\usetikzlibrary{decorations.pathreplacing}
\usetikzlibrary{decorations.pathmorphing}
\usetikzlibrary{decorations.markings}
\usetikzlibrary{arrows.meta}
\usetikzlibrary{positioning}

\usepackage{mathtools}
\usepackage[T1]{fontenc}
\usepackage{tipa}
\usepackage{tipx}
\usepackage{wasysym}
\newcommand{\pthorn}{{\text{\th}}}

\begin{document}

\preprint{APS/123-QED}

\title{Celestial Symmetries of Black Hole Horizons}

\author{Romain Ruzziconi}
\email{romain.ruzziconi@maths.ox.ac.uk}
\affiliation{%
Mathematical Institute, University of Oxford, \\ Andrew Wiles Building, Radcliffe Observatory Quarter, \\
Woodstock Road, Oxford, OX2 6GG, UK
} 

\author{Céline Zwikel}
\email{czwikel@perimeterinstitute.ca}
\affiliation{Perimeter Institute for Theoretical Physics,\\ 31 Caroline Street North, Waterloo, Ontario, Canada N2L 2Y5}

\date{\today}

\begin{abstract}
\noindent We establish a correspondence between the gravitational phase space at null infinity and the subleading phase space near a finite-distance null hypersurface, such as a black hole horizon. Within this framework, we identify the celestial $Lw_{1+\infty}$ symmetries in the subleading phase space at the horizon by constructing their canonical generators and imposing self-duality conditions. This leads to an infinite tower of conserved charges in the absence of radiation, revealing new gravitational observables relevant to black hole physics.
\end{abstract}

\maketitle
\section{Introduction}
Symmetries play a central role in physics, offering a powerful organizing principle. In four-dimensional (4D) gravity with asymptotically flat boundary conditions—relevant for modeling gravitational waves and astrophysical black holes—the symmetry algebra is the BMS algebra \cite{Bondi:1962px, Sachs:1962zza}. This infinite-dimensional extension of Poincaré symmetry includes supertranslations and superrotations \cite{Barnich:2009se, Barnich:2010eb, Campiglia:2014yka, Campiglia:2015yka, Compere:2018ylh}, which place constraints on the infrared structure of scattering amplitudes \cite{Strominger:2017zoo}. The associated charges, constructed via a Noether procedure \cite{Iyer:1994ys, Wald:1999wa, Barnich:2011mi}, encode important dynamical information through flux-balance laws at null infinity, such as the Bondi mass loss formula \cite{Trautman:1958zdi, Bondi:1962px}.

It was early understood that self-dual asymptotically flat spacetimes possess even more symmetries, as revealed by the Penrose non-linear graviton construction in twistor theory \cite{Penrose:1976jq, Penrose:1976js} (see also \cite{Adamo:2021lrv, Mason:2022hly,Bu:2022iak}). In this framework, deformations of the complex structure on twistor space exhibit an $Lw_{1+\infty}$ symmetry, a kind of ``higher-spin generalization'' of BMS symmetry. These symmetries are powerful enough to render the self-dual sector of gravity integrable.

The $Lw_{1+\infty}$ algebra has also emerged in the context of celestial holography, a proposed framework for flat space holography where 4D asymptotically flat gravity is conjectured to be dual to a 2D CFT living on the celestial sphere. Within this setting, the celestial OPEs \cite{Fan:2019emx,Pate:2019lpp,Fotopoulos:2019vac}, derived from the collinear limit of graviton amplitudes, were shown to organize into the $Lw_{1+\infty}$ algebra \cite{Guevara:2021abz, Strominger:2021lvk}. This structure now plays a central role in the formulation of flat space holograms \cite{Costello:2022wso, Costello:2022jpg}.

As in the case of BMS symmetries, it is natural to seek canonical generators for the $Lw_{1+\infty}$ symmetries via a Noether-like construction. Recent progress has achieved this both from a spacetime perspective \cite{Freidel:2021ytz, Geiller:2024bgf}, using the Ashtekar-Streubel radiative phase space at null infinity \cite{Ashtekar:1981bq}, and from a twistor theory approach \cite{Kmec:2024nmu}. These symmetries organize the radial expansion of the metric and are closely related to gravitational multipole moments \cite{Compere:2022zdz}.

A central motivation for studying gravity in asymptotically flat spacetimes, and for developing flat space holography, lies in the desire to understand black holes in the sky \cite{Akiyama:2019cqa}, which may carry essential clues toward a quantum theory of gravity. Black hole symmetries have been shown to account for the Bekenstein-Hawking entropy formula in 3D gravity \cite{Strominger:1997eq}, and this perspective has been extended to 4D Kerr black holes \cite{Guica:2008mu, Castro:2010fd}. Analogues of BMS symmetries at black hole horizons \cite{Hawking:2016msc, Hawking:2016sgy, Haco:2018ske, Donnay:2015abr, Donnay:2016ejv, Grumiller:2019fmp, Adami:2021nnf} — often referred to as ``soft hair'' — have been proposed to play a key role in the microscopic understanding of black hole entropy. However the action of $Lw_{1+\infty}$ symmetries is yet to be unveiled at the horizon and is expected to play a major part in this discussion. 

An important challenge in this endeavor is that the geometry of a null hypersurface at finite distance is much more complicated that the one at null infinity: the induced metric has to be time-dependent to encode radiative degrees of freedom and the dynamics is encoded in Raychaudhuri \cite{PhysRev.98.1123} and Damour \cite{PhysRevD.18.3598,Damour:1979wya} equations. This drastically complicates the phase space discussion, and does not allow to directly import the results from null infinity to the horizon. Relations between null infinity and bulk null hypersurfaces have recently been discussed in \cite{Ashtekar:2024mme,Ashtekar:2024bpi,Riello:2024uvs}. 

The goal of this Letter is to identify the celestial $Lw_{1+\infty}$ symmetries of black hole horizons.\footnote{Most of the discussion in this paper applies to generic null hypersurfaces, not only to black hole horizons. For conciseness, we will slightly abuse the terminology of ``horizons'' throughout the text, as our ultimate goal is to apply this analysis to black hole horizons.} More precisely, we establish two main results:
$(i)$ We demonstrate a precise correspondence between the Ashtekar-Streubel phase space at null infinity and the subleading phase space near a null hypersurface, such as a black hole horizon. This is achieved using Penrose’s conformal compactification, the behavior of the Newman-Penrose (NP) formalism under Weyl rescalings, and covariant phase space techniques. $(ii)$ Leveraging this correspondence, we identify the $Lw_{1+\infty}$ charges at the horizon and show that, upon imposing self-duality conditions, their fluxes integrated over the horizon yield the canonical generators of the celestial symmetries. In particular, in the absence of radiation through the horizon, these charges are conserved, giving rise to an infinite tower of observables relevant for black hole physics.

This Letter is complemented by a companion paper \cite{Ruzziconi:2025fuy}, which provides a comprehensive discussion of the null hypersurface phase space and detailed derivations of some of the results presented here.

\section{NP formalism at null infinity}
\label{sec:NP formalism at null infinity}
In this Section, we review salient features of the phase space at null infinity using NP formalism \cite{Newman:1961qr} (we follow the notations and conventions of \cite{Geiller:2022vto,Geiller:2024bgf}) and list the key ingredients to identify the $Lw_{1+\infty}$ symmetries in that set-up. 

The NP formalism is a powerful framework based on the use of a null tetrad $(\ell,n, m, \bar{m})$ in the first-order Cartan formulation of general relativity: 
\begin{equation}
   g_{\mu\nu} = -\ell_\mu n_\nu - \ell_\nu n_\mu + m_\mu \bar{m}_\nu + m_{\nu} \bar{m}_\mu .
\end{equation}  Notably, this formalism has been applied in asymptotically flat spacetime at null infinity \cite{Newman:1962cia} to derive the solution space and discuss the characteristic initial value problem. The Newman-Unti tetrad has the following expansion near $\mathscr I \equiv \{ r_{\mathscr {I}} \to \infty \}$:
\begin{align} \nonumber
    &\ell = -\partial_{r_\mathscr{I}}, \\ \label{tetrad exapansion}
    &n =(-1+ \mathcal{O}(r^{-1}_{\mathscr I})) \partial_v+\mathcal{O}(r_{\mathscr I}) \partial_{r_\mathscr{I}} + \mathcal{O}(r^{-1}_{\mathscr I})^A \partial_A, \\
    &m = \frac{1}{r_{\mathscr{I}}} m_0^A \partial_A + \mathcal{O}(r_{\mathscr{I}}^{-2}), \quad \bar{m} = \frac{1}{r_{\mathscr{I}}}\bar{m}_0^A \partial_A + \mathcal{O}(r_{\mathscr I}^{-2}) 
    \nonumber
\end{align} where $v$ is the null time along $\mathscr{I}$ and the two-dimensional boundary metric $q_{AB} = m_A^0 \bar{m}_B^0 + \bar m_A^0 m_B^0$ is usually fixed on the phase space. The components of the Weyl tensor expressed in this tetrad are required to obey the Peeling theorem 
\begin{equation}
\label{eq:peeling}
    \Psi_n = \frac{\Psi^0_n}{r^{5-n}_\mathscr{I}} + \mathcal{O} (r^{n-6}_\mathscr{I}) .
\end{equation} In particular the leading component $\Psi_4^0$ is purely radiative. Furthermore, the spin coefficients $\sigma$ and $\lambda$, which play a crucial role in the radiative phase space at $\mathscr{I}$, have the following expansions: 
\begin{equation} \label{expansions spin coefficients}
    \sigma = \frac{\sigma_0}{r^2_{\mathscr{I}}} + \mathcal{O} (r^{-3}_{\mathscr{I}}) , \quad
    \lambda = \lambda_0 + \frac{\lambda_1}{r_{\mathscr{I}}} + \mathcal{O}(r^{-2}_{\mathscr{I}}) .
\end{equation} On-shell, we have $\lambda_0 = \partial_vq_{AB} \bar m_0^A\bar m_0^B= 0$, which is the well-known statement that $\mathscr{I}$ has no intrinsic shear. Additionally, we have $\lambda_1 = \partial_v \bar{\sigma}_0$, which corresponds to the Bondi news \cite{Bondi:1962px}.

Two key ingredients to identify the celestial symmetries at $\mathscr{I}$ from a spacetime perspective \cite{Freidel:2021ytz, Geiller:2024bgf,Kmec:2024nmu} are the following: 
\begin{itemize}

\item The recursion relations encoding the hierarchies of the integrable self-dual sector of gravity are the starting point of the analysis:
    \begin{equation} \label{full recursion at null infinity}
    \begin{split}
         -\partial_v Q_s & = \eth Q_{s-1}-(s+1)\sigma_0Q_{s-2} \\s&= -1, 0, 1, 2, 3, \ldots
         \end{split}
    \end{equation} 
    where $\eth $ encodes angular derivative on the sphere \cite{Eastwood:1982}. Each $Q_s$ for $s>-2$ corresponds to a charge aspect of spin weight $s$ and $Q_{-2} \equiv \Psi^0_4 = \partial_v \lambda_1$ is the radiation responsible for the non-conservation. These recursion relations correspond to an infinite tower of flux-balance laws at null infinity. For $s=-2, \ldots, 2$, we have $Q_{s} = \Psi^0_{2-s}$ and Equation \eqref{full recursion at null infinity} follows from the Bianchi identities. For $s>2$, the $Q_s$'s appear in the $1/r_{\mathscr{I}}$ expansion of $\Psi_0$. Self-dual gravity and full gravity exhibit the same pattern only at leading orders near null infinity. In \cite{Geiller:2024bgf}, deviations with respect to the self-dual sector were observed for $s \ge 4$, and obtaining \eqref{full recursion at null infinity} from the full gravity theory requires a truncation of the Bianchi identities. 
    
\item The Ashtekar-Streubel symplectic structure on the radiative phase space \cite{Ashtekar:1981bq} can be obtained from standard covariant phase space methods \cite{Iyer:1994ys} by pushing a Cauchy slice to $\mathscr{I}$:
\begin{equation} \label{ASv1}
 \boldsymbol{\Omega}_{\mathscr{I}} =\frac1{8\pi G} \int_{\mathscr{I}} dv d^2 x \sqrt{q} \delta \lambda_1 \wedge \delta \sigma_0 +c.c.
\end{equation} where $c.c.$ denotes the complex conjugate terms. The corresponding Poisson bracket is used to compute the $Lw_{1+\infty}$ charge algebra.
    
\end{itemize}

In the next two sections we find the analogue of these two statements at the horizon.

\section{From null infinity to the horizon}
\label{sec:From null infinity to the horizon}
The derivation of the solution space in the NP formalism around a null hypersurface at finite distance will be addressed in full details in \cite{Ruzziconi:2025fuy} (see also \cite{Liu:2022uox} for preliminary results in the non-radiative sector). As mentioned in the introduction, one of the important technical issues is that the intrinsic geometry of the null hypersurface contains genuine gravitational degrees of freedom, by contrast with null infinity. However, as we shall now explain, the use of the NP formalism allows us to drastically simplify the analysis and identify the part of the solution space that is connected to null infinity via Penrose's conformal compactification \cite{Penrose:1962ij, Penrose:1964ge,Newman:1966ub, Penrose:1986uia}. 

One of the advantages of the NP formalism \cite{Newman:1961qr} is to trade tensor fields on spacetime for simpler spin and boost weighted scalars. Additionally, consider the rescaling of the tetrad \cite{Penrose:1984uia}
\begin{equation}
\label{rescaling tetrad}
\begin{split}
    &\ell^\mu \to \Omega^{-2} \ell^\mu, \quad n^\mu \to n^{\mu}, \\
    &m^\mu \to \Omega^{-1} m^\mu, \quad \bar{m}^\mu \to  \Omega^{-1} \bar{m}^\mu
\end{split}
\end{equation} inducing a Weyl rescaling of the metric $g_{\mu\nu} \to \Omega^2 g_{\mu\nu}$, $\Omega> 0$. A weighted scalar $\eta$ has well-defined Weyl weight $W$ if it transforms as $ \eta \to \Omega^W \eta$ under \eqref{rescaling tetrad}. Examples of weighted scalars are given by the components of the Weyl tensors and the spin coefficients $\lambda$ and $\sigma$:
\begin{equation} \label{weights}
    \Psi_n :W = -(5-n), \quad \sigma:W = -2 , \quad \lambda: W = 0 .
\end{equation}
Acting with the bare derivative operators $\partial_v$ and $\eth$ on weighted scalars does not produce well-defined Weyl weighted scalars. To remedy this issue, one can introduce the conformal GHP derivative operators \cite{Geroch:1973am,Penrose:1984uia}:
\begin{equation} \label{substitution}
\begin{split} 
    -\partial_v \to \pthorn_{\mathscr{C}}', \qquad \eth\to \eth_{\mathscr{C}}
\end{split}
\end{equation} constructed from the bare operators by correcting them with spin coefficients, playing the role of a Weyl connection, so that they have well-defined weights. In \cite{Ruzziconi:2025fuy}, we display the explicit expression of these correction terms.

We now have all the tools in hand to map the relevant part of the solution space at null infinity to the one at the horizon. We implement the Penrose conformal compactification \cite{Penrose:1962ij, Penrose:1964ge,Newman:1966ub, Penrose:1986uia} on the (partially) \textit{off-shell} radial expansions presented in Section \ref{sec:NP formalism at null infinity}, as the Einstein equations themselves are not Weyl covariant. Starting from the radial expansion of the tetrad \eqref{tetrad exapansion}, and performing the compactification with conformal factor $\Omega$, see \eqref{rescaling tetrad}, we find 
\begin{align} 
    &\ell \to \Omega^{-2}\partial_{r_\mathscr{I}} = \partial_r , \quad n \to \Omega^0 n=-\partial_v  + \mathcal{O}(r) \label{NU tetrad horizon}\\
    &m \to \Omega^{-1}\Big[\frac{1}{r_{\mathscr{I}}} m^A_0 \partial_A + \mathcal{O}(r_{\mathscr I}^{-1}) \Big] =  m^A_0 \partial_A + \mathcal{O}(r), \nonumber 
\end{align} 
and idem for $\bar{m}$. We identified the conformal factor $\Omega \sim r^{-1}_{\mathscr{I}} \sim r$ with the radial coordinate $r \ge 0$ ($r=0$ is the locus of a finite-distance null hypersurface $\mathcal{H}$), and $v\in \mathbb{R}$ is now the null time along $\mathcal{H}$. By contrast with null infinity, the leading terms $m^A_0$ and $\bar{m}^A_0$ characterizing the intrinsic geometry of $\mathcal{H}$ are now part of the phase space and encode genuine radiative degrees of freedom. Applying the conformal transformation on the Peeling theorem \eqref{eq:peeling}, we find the Taylor expansion
\begin{equation} \label{Taylor expansion}
\begin{split} 
     \Psi_n \to \Psi'_n &= \Omega^{n-5} \left[ \frac{\Psi_n^0}{r_{\mathscr I}^{5-n}} + \mathcal{O}( r_{\mathscr I}^{n-6})) \right] \\
     &= \Psi_n^0 + \mathcal{O}(r)
\end{split}
\end{equation}
where we used the Weyl weights of $\Psi_n$ in \eqref{weights}. By analogy with null infinity, we will show that $\Psi^0_4$ contains information on the radiation going through the horizon. Finally, the spin-coefficient expansions \eqref{expansions spin coefficients} are mapped onto the following expansions at finite distance:
\begin{equation}
    \sigma \to \sigma' = \sigma_0 + \mathcal{O}(r) , \quad \lambda \to \lambda' = \lambda_0 + r \lambda_1 + \mathcal{O}(r^2) .
\end{equation}  A key difference compared to null infinity is that, on-shell, $\lambda_0$ does not generically vanish at finite distance — it coincides with the intrinsic shear of the null hypersurface and we have $\Psi^0_4 =-\pthorn'_{\mathscr C}\lambda_0$. 

In the NP formalism, the tetrad and spin-coefficients equations are not Weyl covariant, but the reursion relations are. Indeed, using the conformal GHP operators \eqref{substitution}, one can ``covariantize'' the recursion relations \eqref{full recursion at null infinity} to obtain 
\begin{equation} \label{recursion relations at the horizon}
 \pthorn'_{\mathscr C} Q_s= \eth_{\mathscr C} Q_{s-1}-(s+1)\sigma_0\, Q_{s-2}
\end{equation} This covariantization is harmless at null infinity, as all the correction terms involved in this procedure vanish at $\mathscr{I}$. Exploiting the Weyl covariance of these expressions, we can perform the conformal rescaling and re-interpret these equations at the horizon. However, at finite distance, the correction terms in the GHP derivative operators with respect to the bare operators are crucial as the intrinsic geometry at the horizon is non-trivial. For example, one of these corrections involves the surface gravity of the black hole. Hence, \eqref{recursion relations at the horizon} constitutes the recursion relations at the horizon and, in particular, $Q_s = \Psi_{2-s}^0$, $s=-2,\ldots,2$ are now the components of the Weyl tensor in the Taylor expansion \eqref{Taylor expansion}, and the $Q_s$'s for $s> 2$ appear in the expansion of $\Psi_0$. Notice that, for spin $s=-1,0,1,2$, \eqref{recursion relations at the horizon} are the Bianchi identities at finite distance.

\section{Subleading phase space}
As mentioned in Section \ref{sec:NP formalism at null infinity}, a key ingredient to identify the $Lw_{1+\infty}$ symmetries from a spacetime perspective at null infinity are self-duality conditions. These are necessary only at subleading order in the radial expansion of the spin coefficients and recursion relations, where the deviation between self-dual and full gravity starts to occur \cite{Freidel:2021ytz, Geiller:2024bgf}. At finite distance, self-duality conditions are already visible at leading order in $r$ since the intrinsic geometry on the null hypersurface contains genuine degrees of freedom involving both helicities. We propose conditions on the spin coefficients:
\begin{equation}\label{dualcond1}
    \mu_0=0\,, \quad \bar\lambda_0=0\,, \quad 
\text{Re}(\gamma_0)=\text{constant}
\end{equation} where $\mu_0$ is the expansion of $\mathcal H$ and $\text{Re}(\gamma_0)=-2\kappa$ with $\kappa$ the surface gravity. Here we work in the usual complexified set-up where the functions and their barred versions are complex and independent (e.g. $\lambda_0$ and $\bar{\lambda}_0$ are not complex conjugate). The conditions \eqref{dualcond1} ensure that $[ \pthorn'_{\mathscr{C}}, \eth_{\mathscr{C}}  ] =0$. Furthermore, we will require the following constraints on the variations:
\begin{equation}
  \delta m_A^0=0 \,,\quad \delta \text{Re}(\gamma_0)=0  \,, \quad\delta \sqrt{q}=0 , \quad \delta \tau_0 = 0
  \label{dualcond2}
\end{equation}
where $\tau_0$ is related to the twist of $\mathcal H$. 
One can check that the above conditions \eqref{dualcond1} and \eqref{dualcond2} are consistent with each other through the NP equations for the spin coefficients \cite{Newman:1961qr}, and imply self-duality conditions eliminating one of the two graviton helicities. Generically, $\Psi_4^0 = -\pthorn'_{\mathscr C}\lambda_0 \neq 0$, so that there is still one helicity radiating through the horizon.  

The Einstein-Hilbert symplectic structure $\boldsymbol{\Omega}$ can be obtained from the standard covariant phase space methods \cite{Iyer:1994ys}. For the radial expansion near a null hypersurface obtained in Section \ref{sec:From null infinity to the horizon}, we have
\begin{equation}
    \boldsymbol{\Omega} = \boldsymbol{\Omega}^{(0)} + r \boldsymbol{\Omega}^{(1)} + \mathcal{O}(r^2)
\end{equation} where $\boldsymbol{\Omega}^{(0)}$ is the usual leading symplectic structure discussed in previous literature, see e.g. \cite{Adami:2021nnf}, and $\boldsymbol{\Omega}^{(1)}$ is the subleading symplectic structure. Remarkably, under the conditions \eqref{dualcond1} and \eqref{dualcond2}, the leading symplectic structure vanishes, $\boldsymbol{\Omega}^{(0)} = 0$, and we find
\begin{equation}\label{subleading symplectic structure}
  \boldsymbol{\Omega} = \frac{r}{8\pi G}\,  \oint \int dv \,\delta \sigma_0 \wedge \delta \lambda_0 + \delta \bar{p}^A \wedge \delta \bar{m}^0_A +\mathcal{O}(r^2)     
\end{equation}
where $ \oint = \int d^2x \sqrt{q}$.
 The first term is the analogue of the Ashtekar-Steubel symplectic structure \eqref{ASv1} at infinity for one helicity sector (i.e. there is no $c.c.$ term). An important difference compared to the canonical pair at $\mathscr{I}$ is that, on-shell, $\lambda_0 =-\bar m^A_0\pthorn'_{\mathscr C} \bar m_A^0$ is the intrinsic shear of $\mathcal{H}$ and is not related to the extrinsic shear $\sigma_0$. The bracket between these symplectic variables can be deduced by inverting the symplectic structure 
\begin{equation} \label{canonical bracket}
  \{ \sigma_0 (v_1, x^A_1) , \lambda_0 (v_2, x^A_2) \} = \frac{8 \pi G}{r \sqrt{q}}  \delta^2 (x^A_1 - x^A_2) \delta (v_1-v_2) .
\end{equation}
The second symplectic pair in \eqref{subleading symplectic structure} involves $\bar p^A$, whose explicit expression in terms of the spin coefficients will be provided in \cite{Ruzziconi:2025fuy}, and $\bar m^0_A$. As we will see in the next section, the latter canonical variable does not appear in the $Lw_{1+\infty}$ symmetry generators. Hence, it will not play any role in the computation of the $Lw_{1+\infty}$ charge algebra.

This concludes the identification between the leading Ashtekar-Streubel phase space at $\mathscr{I}$ and the subleading phase space at the horizon, which is the first main result of this Letter. We will exploit this direct connection in the next section to identify the celestial symmetries of the horizon.

\section{Canonical $Lw_{1+\infty}$ symmetries}
The construction of the $Lw_{1+\infty}$ surface charges $H_s$ at the horizon follows a similar heuristic procedure than the one used at null infinity \cite{Freidel:2021ytz,Geiller:2024bgf}, and we refer to \cite{Ruzziconi:2025fuy} for details. We require the following two criteria: $(i)$ the charges are conserved in the absence of radiation characterized by $\Psi^0_4 = 0$, and $(ii)$ their associated integrated fluxes are the canonical generators of $Lw_{1+\infty}$ symmetries in the subleading phase space at the horizon. The general form of these $Lw_{1+\infty}$ surface charges, defined on a cut $v=\text{constant}$ of a $r= \text{constant}$ hypersurface close to the horizon is
\begin{equation} \label{surface charges}
    H_s=\frac{r}{8\pi G}\oint T_{s}Q_{s} + \text{correction terms} .
\end{equation}
The charges are labeled by $s=-1, 0, 1, \ldots$, the spin weight of their associated charge aspect $Q_s$. The $T_s$'s are the associated symmetry parameters. These are weighted scalars satisfying $\pthorn'_{\mathscr{C}} T_s = 0$, generalizing the time-independence condition of the parameters at null infinity. The global factor $r$ indicates that these charges are associated with the subleading phase space at the horizon \eqref{subleading symplectic structure}. The correction terms in \eqref{surface charges} are determined by requiring the criterium of conservation, $(i)$. The explicit form of these correction terms can be found in \cite{Ruzziconi:2025fuy} up to spin $2$. Using the recursion relations \eqref{recursion relations at the horizon}, the commutation properties of the GHP operators \eqref{substitution} (see \cite{Geroch:1973am,Penrose:1984uia}), and the self-duality conditions \eqref{dualcond1}, we show that 
\begin{equation}
    \Psi^0_4 \equiv Q_{-2}= 0 \quad \Longrightarrow \quad \frac{d H_s}{dv} = 0
\end{equation} for all $s \ge -1$. $\Psi^0_4$ is responsible for the non-conservation of the charges, and can thus be identified with the radiation through the horizon. In the absence of it, we have an infinite tower of conserved charges.

Analogously to the situation at null infinity, we can show that the symmetries $s=0$ and $s=1$ are related to the supertranslation and superrotations symmetries at the horizon \cite{Donnay:2015abr, Donnay:2016ejv, Grumiller:2019fmp} and refer to \cite{Ruzziconi:2025fuy} for details. However, our charges are associated with the subleading phase space and are thus of different nature than those considered in previous literature. Furthermore, the higher spin charges provide an infinite tower of new symmetries and observables that do not have a direct diffeomorphism interpretation. For the Kerr solution, we have $H_{-1}^{\text{Kerr}} = 0$, and $H^{\text{Kerr}}_0$ encodes the mass parameter $m$ while $H^{\text{Kerr}}_1$ probes its angular momentum parameter $a$. $H^{\text{Kerr}}_2$ is non vanishing and involves a combination of $m$ and $a$. We expect these charges to be interesting observables for an observer sitting just outside of the horizon.

The integrated fluxes are defined by
\begin{equation}
    \mathcal{F}_s = \int dv \partial_v H_s = H_s|_{v=+\infty} - H_s|_{v=-\infty} \label{integrated fluxes}
\end{equation} and their explicit expression appear in \cite{Ruzziconi:2025fuy} up to spin $2$. The integral over $v$ converges for all $s$ if, as in \cite{Freidel:2021ytz, Geiller:2024bgf, Freidel:2022skz, Kmec:2024nmu}, we impose Schwartzian falloffs: $\lim_{v \to \pm \infty}\lambda_0 \sim e^{-|v|^2}$. This is consistent with the action of the $Lw_{1+\infty}$ symmetries on the phase space if the parameters satisfy the wedge condition: $\eth_{\mathscr{C}}^{s+2} T_s = 0$. Under this assumption, the soft terms in the fluxes vanish and, in particular, $\mathcal{F}_{-1} = 0$.

Finally, using \eqref{canonical bracket} and imposing the self-duality conditions \eqref{dualcond1} and \eqref{dualcond2}, the computation of the algebra follows similar steps than those performed at null infinity \cite{Freidel:2021ytz,Geiller:2024bgf} (see also \cite{Cresto:2024fhd,Cresto:2024mne,Kmec:2024nmu} for non-perturbative results). We find 
\begin{equation}  \label{representation of Linfinity}
\begin{split}
   &\{ \mathcal{F}_{T_{s_1}}, \mathcal{F}_{T_{s_2}} \} = \mathcal{F}_{T_{s_1+ s_2 - 1}}, \\
   &T_{s_1+ s_2 - 1} = ({s_2} + 1) T_{s_2}\eth_{\mathscr{C}} T_{s_1} - ({s_1} + 1)T_{s_1} \eth_{\mathscr{C}} T_{s_2} 
\end{split}
\end{equation} 
where we noted $\mathcal{ F}_s \equiv \mathcal{F}_{T_s}$. This corresponds to a representation of the $Lw_{1+\infty}$ algebra in terms of weighted scalars, hence satisfying the requirement $(ii)$. Therefore, we have shown that $\mathcal{F}_s$ are the canonical generators for the $Lw_{1+\infty}$ symmetries at the horizon.

\section{Discussion}
In this Letter, we have established a map between the Ashtekar-Streubel phase space at null infinity and the subleading phase space at the horizon. Using this correspondence, we have imported the celestial symmetries from null infinity to the horizon. This work contributes to relating physics at null infinity and at finite distance.

Indeed, our work provides a clear set-up to apply
the ideas of celestial holography to black hole physics. For instance, it would be rewarding to relate our analysis with \cite{Crawley:2023brz} where a holographic state for a self-dual black hole is provided, and understand the precise role of the subleading phase space in this context. 

The celestial $Lw_{1+\infty}$ symmetries at null infinity possess a natural interpretation on twistor space. It would be interesting to derive the surface charge expressions obtained here from first principles using similar method as in \cite{Kmec:2024nmu}. This would require defining an analogue of asymptotic twistor space adapted to black hole horizons.

As mentioned in the introduction, from the point of view of the horizon, symmetries corresponding to soft hair could account for black hole entropy. It would be interesting to study whether the $Lw_{1+\infty}$ symmetries play a role in this discussion. 

Finally, our construction provides meaningful observables for an observer outside of the black hole, but close enough to probe the first subleading order of the metric. It would be interesting to understand the significance of the $Lw_{1+\infty}$ charges in the recent developments of black hole physics, such as the analysis of the photon ring \cite{Bardeen_1974,Luminet,Akiyama:2019cqa,Johnson:2019ljv} or the use of multipole moments at the horizon in numerical relativity \cite{Ashtekar:2004gp,Schnetter:2006yt,Pook-Kolb:2020jlr,Ashtekar:2021wld}.

\begin{acknowledgments}
It is our pleasure to thank Nicolas Cresto, Laurent Freidel, Marc Geiller, Adam Kmec and Lionel Mason for useful discussions. RR is
supported by the Titchmarsh Research Fellowship at the Mathematical Institute and by
the Walker Early Career Fellowship at Balliol College. RR also thanks the Perimeter Institute for its hospitality during the Celestial Holography Summer School 2024 for the Simons Collaboration, where an important part of this work was completed. Research at Perimeter Institute is supported in part by the Government of Canada through the Department of Innovation, Science and Economic Development Canada and by the Province of Ontario through the Ministry of Colleges and Universities. 

\end{acknowledgments}

\bibliographystyle{style}
\bibliography{biblio}

\end{document}